\documentclass[11pt]{article}
\usepackage{epstopdf}
\usepackage{color}
\usepackage{subfigure}
\usepackage{amsmath}
\usepackage{amssymb}
\usepackage{graphicx,color}
\usepackage{cite}
\usepackage{enumerate}
\usepackage{amsthm}
\usepackage{amsfonts,mathrsfs}
\usepackage{geometry}
\usepackage{ifthen}

\begin{document}

\title{Uncertainty relations based on modified Wigner-Yanase-Dyson skew information}

\vskip0.1in
\author{\small Zhaoqi Wu$^{1,5}$, Lin
Zhang$^{2,5}$\thanks{Corresponding author. E-mail: godyalin@163.com;linzhang@mis.mpg.de}, Jianhui Wang$^{3}$, Xianqing Li-Jost$^{5}$, Shao-Ming Fei$^{4,5}$\thanks{Corresponding author. E-mail: feishm@cnu.edu.cn}\\
{\small\it  1. Department of Mathematics, Nanchang University, Nanchang 330031, P R China} \\
{\small\it  2. Institute of Mathematics, Hangzhou Dianzi University, Hangzhou 310018, P R China}\\
{\small\it  3. Department of Physics, Nanchang University, Nanchang 330031, P R China}\\
{\small\it  4. School of Mathematical Sciences, Capital Normal
University, Beijing 100048, P R China}\\
{\small\it  5. Max-Planck-Institute for Mathematics in the Sciences,
04103 Leipzig, Germany}}
\date{}
\maketitle

\noindent {\bf Abstract} {\small } Uncertainty relation is a core issue in quantum mechanics and quantum information theory. We introduce modified generalized Wigner-Yanase-Dyson (MGWYD) skew information and modified weighted generalized Wigner-Yanase-Dyson (MWGWYD) skew information, and establish new uncertainty relations in terms of the MGWYD skew information and MWGWYD skew information.

\vskip0.2in

\noindent {\bf 1. Introduction}

\vskip0.1in

Let $H$ be a separable complex Hilbert space and $B(H)$, $S(H)$ and
$D(H)$ the set of all bounded linear operators, Hermitian operators
and density operators (positive operators of trace $1$) on $H$,
respectively. An operator $A\in B(H)$ is called a {\it trace-class}
operator if
$$\|A\|_{1}:=\sum_{n\in I}\langle e_n||A||e_n\rangle<\infty$$
for some orthonormal basis $\{e_n\}_{n\in I}$ for $H$, where
$|A|=(A^\dag A)^{\frac{1}{2}}$. In this case, the {\it trace} of $A$
is defined as $\mathrm{Tr}(A)=\sum_{n\in I}\langle
e_n|A|e_n\rangle$. We denote by $L^1(H)$ the set of all trace-class
operators on $H$. An operator $A\in B(H)$ is called a {\it
Hilbert-Schmidt} operator if
$$\|A\|_{2}:=(\sum_{n\in I}\langle e_n|A^\dag A|e_n\rangle)^{\frac{1}{2}}<\infty$$
for some orthonormal basis $\{e_n\}_{n\in I}$ for $H$. We denote by
$L^2(H)$ the set of all Hilbert-Schmidt operators on $H$.

For a density operator $\rho\in D(H)$ and an observable $A\in S(H)$,
the {\it Wigner-Yanase} (WY) skew information \cite{WY} was defined
by
\begin{equation}\label{eq1}
\mathrm{I}_{\rho}(A):=-\frac{1}{2}\mathrm{Tr}([\rho^{\frac{1}{2}},A]^2),
\end{equation}
where $[X,Y]:=XY-YX$ is the commutator of $X$ and $Y$. A more
general quantity was suggested by Dyson,
\begin{equation}\label{eq2}
\mathrm{I}_{\rho}^{\alpha}(A):=-\frac{1}{2}\mathrm{Tr}([\rho^{\alpha},A][\rho^{1-\alpha},A]),\,\,~0\leq
\alpha \leq 1,
\end{equation}
which is now called the {\it Wigner-Yanase-Dyson} (WYD) skew
information. The quantity in Eq. (\ref{eq2}) was further generalized
in \cite{CL} to
\begin{equation}\label{eq3}
\mathrm{I}_{\rho}^{\alpha,\beta}(A)=-\frac{1}{2}\mathrm{Tr}([\rho^\alpha,
A][\rho^\beta, A]\rho^{1-\alpha-\beta}),~~~\alpha,\beta\geq
0,~\alpha+\beta\leq 1,
\end{equation}
which is termed the {\it generalized Wigner-Yanase-Dyson} (GWYD) skew information.
It is easy to see that when $\alpha+\beta=1$, Eq. (\ref{eq3}) reduces to Eq. (\ref{eq2}), and Eq. (\ref{eq2})
reduces to Eq. (\ref{eq1}) when $\alpha=\frac{1}{2}$.

Another generalization of WYD skew information was defined in
\cite{FURU1} as
\begin{equation}\label{eq4}
\mathrm{K}_{\rho}^{\alpha}(A)=-\frac{1}{2}\mathrm{Tr}\left(\left[\frac{\rho^\alpha+\rho^{1-\alpha}}{2},
A_0\right]^2\right) ~,\,\,0\leq \alpha \leq 1,
\end{equation}
where $A_0=A-\mathrm{Tr}(\rho A)I$. We call $\mathrm{K}_{\rho}^{\alpha}(A)$ the {\it weighted
Wigner-Yanase-Dyson skew information} in the following. Noting that
$\mathrm{I}_{\rho}(A)=\mathrm{I}_{\rho}(A_0)$, when
$\alpha=\frac{1}{2}$, Eq. (\ref{eq4}) also reduces to Eq.
\eqref{eq1} in this case.

Remarkable properties of WYD skew information and GWYD skew
information have been revealed, and various types of uncertainty
relations based on WY skew information, WYD skew information and
GWYD skew information have been studied during the past few years
\cite{LUO1,LUO2,LUO3,LUO4,LUO5,LUO6,LUO7,LUO8,LUO9,LUO10,FURU2,FURU3,YANA1,YANA2,KO,CB1}.
Particularly, uncertainty relations based on WY skew information and
WYD skew information with quantum memory have been investigated
recently \cite{MCF,LF}. Besides skew information, uncertainty
relations based on other quantities such as entropy, variance,
statistical distance and quantum coherence have been extensively
studied with experimental demonstrations
\cite{XYL,CB2,LUO11,RAE,QHH,MWC1,MWC2,MYL}. It is well known that
the observables and Hamiltonians in quantum mechanics are assumed to
be Hermitian operators mathematically. However, it is argued that
non-Hermitian quantum mechanics may also be an interesting framework
\cite{MN}. Moreover, other important operators such as quantum gates
\cite{NC}, generalized quantum gates \cite{LGL} and the Kraus
operators of a quantum channel \cite{NC} are not necessarily
Hermitian. Therefore, it is natural to consider the corresponding
definitions of the different types of the skew information mentioned
above for pseudo-Hermitian and/or {\cal PT}-symmetric quantum
mechanics
\cite{Bender98,Makris-PRL,Guo-PRL,Ruter-NP,Chang-NP,Tang-NP}.

For a density operator $\rho\in D(H)$ and an operator $A\in L^2(H)$
(not necessarily Hermitian), a generalization of the quantity in Eq.
(\ref{eq1}) is defined by \cite{DOU1}
\begin{equation}\label{eq5}
|\mathrm{I}_{\rho}|(A):=-\frac{1}{2}\mathrm{Tr}([\rho^{\frac{1}{2}},A^\dag][\rho^{\frac{1}{2}},A]),
\end{equation}
which we refer to {\it modified Wigner-Yanase} (MWY) skew information.

Similarly, a generalization of the quantity in Eq. (\ref{eq2}) is defined by \cite{DOU2}
\begin{equation}\label{eq6}
|\mathrm{I}_{\rho}^{\alpha}|(A):=-\frac{1}{2}\mathrm{Tr}([\rho^{\alpha},A^\dag][\rho^{1-\alpha},A]),\,\,~0\leq
\alpha \leq 1,
\end{equation}
for any $A\in L^2(H)$ and $\rho\in D(H)$, which we call {\it modified Wigner-Yanase-Dyson} (MWYD) skew information.

And a generalization of the quantity in Eq. (\ref{eq4}) is given by \cite{CZL2}
\begin{equation}\label{eq7}
|\mathrm{K}_{\rho}^{\alpha}|(A)=-\frac{1}{2}\mathrm{Tr}\left(\left[\frac{\rho^\alpha+\rho^{1-\alpha}}{2},
A_0^\dag\right]\left[\frac{\rho^\alpha+\rho^{1-\alpha}}{2},
A_0\right]\right) ,\,\,~0\leq \alpha \leq 1,
\end{equation}
for any $A\in L^2(H)$ and $\rho\in D(H)$, which we call {\it modified weighted Wigner-Yanase-Dyson} (MWWYD) skew information.

In \cite{DOU1}, the authors established a Heisenberg type
uncertainty relation and a Schr\"odinger-type uncertainty relation
based on MWY skew information. The definitions and properties of
MWYD skew information were discussed in \cite{DOU2}, and the
uncertainty relations for MWY skew information and MWYD skew
information were extensively studied in \cite{LI,CZL1}. Moreover,
the uncertainty relations for MWWYD skew information were given in
\cite{CZL2}. Recently, the authors in \cite{FYJ} introduced some
related quantities, and derived some generalizations of
Schr\"odinger's uncertainty and Heisenberg uncertainty relations
described by MWYD skew information.

In this paper, we first introduce the concepts of modified
generalized Wigner-Yanase-Dyson (MGWYD) skew information and
modified weighted generalized Wigner-Yanase-Dyson (MWGWYD) skew
information as the generalizations of the quantities defined in Eqs.
(\ref{eq3}) and (\ref{eq7}), and discuss their properties in detail
in Section 2. Furthermore, we provide new uncertainty relations
based on these two new quantities in Section 3. Some concluding
remarks are given in Section 4.

\vskip0.1in
\noindent {\bf 2. MGWYD and MWGWYD skew information}

\vskip0.1in

We first define the MGWYD skew information for an operator $A\in L^2(H)$ (not necessarily Hermitian) and $\rho\in D(H)$ as follows:
\begin{equation}\label{eq8}
|\mathrm{I}_{\rho}^{\alpha,\beta}|(A)=-\frac{1}{2}\mathrm{Tr}([\rho^\alpha,
A^\dag][\rho^\beta, A]\rho^{1-\alpha-\beta}),~~\alpha,\beta\geq
0,~\alpha+\beta\leq 1.
\end{equation}

Correspondingly we define
\begin{equation}\label{eq9}
|\mathrm{J}_{\rho}^{\alpha,\beta}|(A)=\frac{1}{2}\mathrm{Tr}(\{\rho^\alpha,
A_0^\dag]\}\{\rho^\beta,
A_0\}\rho^{1-\alpha-\beta}),~~\alpha,\beta\geq 0,~\alpha+\beta\leq
1.
\end{equation}
where $\{X,Y\}:=XY+YX$ is the anti-commutator of $X$ and $Y$. It
follows from the definitions that
$$|\mathrm{I}_{\rho}^{\alpha,\beta}|(A)=\frac{1}{2}[\mathrm{Tr}(\rho A^\dag A)+\mathrm{Tr}(\rho^{\alpha+\beta}A\rho^{1-\alpha-\beta}A^\dag)-\mathrm{Tr}(\rho^{1-\beta}A^\dag \rho^\beta A)-\mathrm{Tr}(\rho^\alpha A \rho^{1-\alpha}A^\dag)]$$
and
$$|\mathrm{J}_{\rho}^{\alpha,\beta}|(A)=\frac{1}{2}[\mathrm{Tr}(\rho A^\dag A)+\mathrm{Tr}(\rho^{\alpha+\beta}A\rho^{1-\alpha-\beta}A^\dag)+\mathrm{Tr}(\rho^{1-\beta}A^\dag \rho^\beta A)+\mathrm{Tr}(\rho^\alpha A \rho^{1-\alpha}A^\dag)].$$

We also need the following definitions of related quantities.

{\bf Definition 1} For $\alpha,\beta\geq 0$, $\alpha+\beta\leq 1$, $A,B\in L^2(H)$ and $\rho\in D(H)$, we define the following quantities:

(i)
$|\mathrm{Cov}_{\rho}^{\alpha,\beta}|(A,B)=\frac{1}{2}[\mathrm{Tr}(\rho
A^\dag
B)+\mathrm{Tr}(\rho^{\alpha+\beta}A\rho^{1-\alpha-\beta}B^\dag)-\mathrm{Tr}(\rho
B)Tr(\rho A^\dag)-\mathrm{Tr}(\rho A)Tr(\rho B^\dag)]$;

(ii)
$|\mathrm{Var}_{\rho}^{\alpha,\beta}|(A)=\frac{1}{2}[\mathrm{Tr}(\rho
A^\dag
A)+\mathrm{Tr}(\rho^{\alpha+\beta}A\rho^{1-\alpha-\beta}A^\dag)-\mathrm{Tr}(\rho
A)\mathrm{Tr}(\rho A^\dag)-\mathrm{Tr}(\rho A)\mathrm{Tr}(\rho
A^\dag)]$;

(iii)
$|\mathrm{Corr}_{\rho}^{\alpha,\beta}|(A,B)=\frac{1}{2}[\mathrm{Tr}(\rho
A^\dag
B)+\mathrm{Tr}(\rho^{\alpha+\beta}A\rho^{1-\alpha-\beta}B^\dag)-\mathrm{Tr}(\rho^{1-\beta}A^\dag
\rho^\beta B)-\mathrm{Tr}(\rho^\alpha A \rho^{1-\alpha}B^\dag)]$;

(iv)
$|\mathrm{C}_{\rho}^{\alpha,\beta}|(A,B)=\frac{1}{2}[\mathrm{Tr}(\rho^{1-\beta}A^\dag
\rho^\beta B)+\mathrm{Tr}(\rho^\alpha A \rho^{1-\alpha}B^\dag)]$;

(v)
$|\mathrm{C}_{\rho}^{\alpha,\beta}|(A)=|\mathrm{C}_{\rho}^{\alpha,\beta}|(A,A)=\frac{1}{2}[\mathrm{Tr}(\rho^{1-\beta}A^\dag
\rho^\beta A)+\mathrm{Tr}(\rho^\alpha A \rho^{1-\alpha}A^\dag)]$;

(vi)
$|\mathrm{U}_{\rho}^{\alpha,\beta}|(A)=\sqrt{|\mathrm{Var}_{\rho}^{\alpha,\beta}|(A)^2-[|\mathrm{Var}_{\rho}^{\alpha,\beta}|(A)-|\mathrm{I}_{\rho}^{\alpha,\beta}|(A)]^2}$.

The following proposition follows immediately from the above definitions.

{\bf Proposition 1} For $\alpha,\beta\geq 0$, $\alpha+\beta\leq 1$, $A\in L^2(H)$ and $\rho\in D(H)$, it holds that

(i)
$|\mathrm{I}_{\rho}^{\alpha,\beta}|(A)=|\mathrm{I}_{\rho}^{\beta,\alpha}|(A)$,
$|\mathrm{J}_{\rho}^{\alpha,\beta}|(A)=|\mathrm{J}_{\rho}^{\beta,\alpha}|(A)$;

(ii)
$|\mathrm{I}_{\rho}^{\alpha,\beta}|(A^\dag)=|\mathrm{I}_{\rho}^{\alpha,\beta}|(A)$,
$|\mathrm{J}_{\rho}^{\alpha,\beta}|(A^\dag)=|\mathrm{J}_{\rho}^{\alpha,\beta}|(A)$;

(iii)
$|\mathrm{I}_{\rho}^{\alpha,\beta}|(A)=|\mathrm{Corr}_{\rho}^{\alpha,\beta}|(A,A)$.

In order to obtain the main results in the next section, we first study the properties of the MGWYD skew information and the related quantities defined above.

{\bf Proposition 2} Let $\alpha,\beta\geq 0$, $\alpha+\beta\leq 1$, $A,B\in L^2(H)$, $\rho\in D(H)$, $A_0=A-\mathrm{Tr}(\rho A)I$, $B_0=B-\mathrm{Tr}(\rho B)I$. We have

(i)
$|\mathrm{Corr}_{\rho}^{\alpha,\beta}|(A,B)=|\mathrm{Corr}_{\rho}^{\alpha,\beta}|(A_0,B_0)=|\mathrm{Cov}_{\rho}^{\alpha,\beta}|(A,B)-|\mathrm{C}_{\rho}^{\alpha,\beta}|(A_0,B_0)$;

(ii)
$|\mathrm{I}_{\rho}^{\alpha,\beta}|(A)=|\mathrm{I}_{\rho}^{\alpha,\beta}|(A_0)=|\mathrm{Var}_{\rho}^{\alpha,\beta}|(A)-|\mathrm{C}_{\rho}^{\alpha,\beta}|(A_0)=
\frac{1}{2}[\mathrm{Tr}(\rho A^\dag
A)+\mathrm{Tr}(\rho^{\alpha+\beta}A\rho^{1-\alpha-\beta}A^\dag)-\mathrm{Tr}(\rho^{1-\beta}A^\dag
\rho^\beta A)-\mathrm{Tr}(\rho^\alpha A \rho^{1-\alpha}A^\dag)]$;

(iii)
$|\mathrm{J}_{\rho}^{\alpha,\beta}|(A)=|\mathrm{Var}_{\rho}^{\alpha,\beta}|(A)+|\mathrm{C}_{\rho}^{\alpha,\beta}|(A_0)=
\frac{1}{2}[\mathrm{Tr}(\rho A^\dag
A)+\mathrm{Tr}(\rho^{\alpha+\beta}A\rho^{1-\alpha-\beta}A^\dag)+\mathrm{Tr}(\rho^{1-\beta}A^\dag
\rho^\beta A)+\mathrm{Tr}(\rho^\alpha A \rho^{1-\alpha}A^\dag)]$;

(iv)
$|\mathrm{U}_{\rho}^{\alpha,\beta}|(A)=\sqrt{|\mathrm{I}_{\rho}^{\alpha,\beta}|(A)|\mathrm{J}_{\rho}^{\alpha,\beta}|(A)}$;

(v) $0\leq |\mathrm{I}_{\rho}^{\alpha,\beta}|(A)\leq
|\mathrm{U}_{\rho}^{\alpha,\beta}|(A)\leq
|\mathrm{Var}_{\rho}^{\alpha,\beta}|(A)$.

{\bf Proof.} We first prove (i). It is direct to check that
\begin{eqnarray*}
&&|\mathrm{Corr}_{\rho}^{\alpha,\beta}|(A_0,B_0)
\\&=&\frac{1}{2}[\mathrm{Tr}(\rho(A-\mathrm{Tr}(\rho A)I)^\dag B)+\mathrm{Tr}(\rho^{\alpha+\beta}(A-\mathrm{Tr}(\rho A)I)\rho^{1-\alpha-\beta}B^\dag)\notag
\\&&-\mathrm{Tr}(\rho^{1-\beta}(A-\mathrm{Tr}(\rho A)I)^\dag \rho^\beta B)-\mathrm{Tr}(\rho^\alpha(A-\mathrm{Tr}(\rho A)I)\rho^{1-\alpha}B^\dag)]
\\&=&\frac{1}{2}[\mathrm{Tr}(\rho(A-\mathrm{Tr}(\rho A)I)^\dag B)+\mathrm{Tr}(\rho^{\alpha+\beta}(A-\mathrm{Tr}(\rho A)I)\rho^{1-\alpha-\beta}B^\dag)\notag
\\&&-\mathrm{Tr}(\rho^{1-\beta}(A-\mathrm{Tr}(\rho A)I)^\dag \rho^\beta B)-\mathrm{Tr}(\rho^\alpha(A-\mathrm{Tr}(\rho A)I)\rho^{1-\alpha}B^\dag)]
\\&=&\frac{1}{2}[(\mathrm{Tr}(\rho A^\dag B)-\mathrm{Tr}(\rho B)\mathrm{Tr}(\rho A^\dag)-\overline{\mathrm{Tr}(\rho A)}\mathrm{Tr}(\rho B)+\overline{\mathrm{Tr}(\rho A)}\mathrm{Tr}(\rho B))\notag
\\&&+(\mathrm{Tr}(\rho^{\alpha+\beta}A\rho^{1-\alpha-\beta}B^\dag)-\overline{\mathrm{Tr}(\rho B)}\mathrm{Tr}(\rho A)-\mathrm{Tr}(\rho A)\mathrm{Tr}(\rho B^\dag)+\mathrm{Tr}(\rho A)\overline{\mathrm{Tr}(\rho B)})\notag
\\&&-(\mathrm{Tr}(\rho^{1-\beta}A^\dag \rho^\beta B)-\mathrm{Tr}(\rho B)\mathrm{Tr}(\rho A^\dag)-\overline{\mathrm{Tr}(\rho A)}\mathrm{Tr}(\rho B)+\overline{\mathrm{Tr}(\rho A)}\mathrm{Tr}(\rho B))\notag
\\&&-(\mathrm{Tr}(\rho^\alpha A \rho^{1-\alpha}B^\dag)-\overline{\mathrm{Tr}(\rho B)}\mathrm{Tr}(\rho A)-\mathrm{Tr}(\rho A)\mathrm{Tr}(\rho B^\dag)+\mathrm{Tr}(\rho A)\overline{\mathrm{Tr}(\rho B)})]
\\&=&\frac{1}{2}[\mathrm{Tr}(\rho A^\dag B)+\mathrm{Tr}(\rho^{\alpha+\beta}A\rho^{1-\alpha-\beta}B^\dag)-\mathrm{Tr}(\rho^{1-\beta}A^\dag \rho^\beta B)-\mathrm{Tr}(\rho^\alpha A \rho^{1-\alpha}B^\dag)]
\\&=&|\mathrm{Corr}_{\rho}^{\alpha,\beta}|(A,B).
\end{eqnarray*}
Meanwhile we have
\begin{eqnarray*}
|\mathrm{Corr}_{\rho}^{\alpha,\beta}|(A_0,B_0)
&=&\frac{1}{2}[\mathrm{Tr}(\rho A_0^\dag
B_0)+\mathrm{Tr}(\rho^{\alpha+\beta}A_0\rho^{1-\alpha-\beta}B_0^\dag)]-|\mathrm{C}_{\rho}^{\alpha,\beta}|(A_0,B_0)
\\&=&|\mathrm{Cov}_{\rho}^{\alpha,\beta}|(A,B)-|\mathrm{C}_{\rho}^{\alpha,\beta}|(A_0,B_0).
\end{eqnarray*}
Hence, (i) holds. Then (ii) can be easily obtained. (iii) can be proved analogously. Similar to the proof of Theorem 2 in \cite{FYJ}, (iv) and (v) can be deduced similarly.
$\Box$

{\bf Proposition 3} Let $\alpha,\beta\geq 0$, $\alpha+\beta\leq 1$, $A,B\in L^2(H)$, $\rho\in D(H)$, $A_0=A-\mathrm{Tr}(\rho A)I$ and $ B_0=B-\mathrm{Tr}(\rho B)I$. For a spectral decomposition of $\rho=\sum_{m}\lambda_m |\psi_m\rangle\langle \psi_m|$, denote $a_{mn}=\langle\psi_m|A_0|\psi_n \rangle$. We have

(i) $|\mathrm{Cov}_{\rho}^{\alpha,\beta}|(A,B)
=\frac{1}{2}\sum_{mn}\lambda_m^{\alpha+\beta}(\lambda_m^{1-\alpha-\beta}\overline{a_{nm}}b_{nm}+\lambda_n^{1-\alpha-\beta}a_{mn}\overline{b_{mn}})$.
In particular, $|\mathrm{Var}_{\rho}^{\alpha,\beta}|(A)
=\frac{1}{2}\sum_{mn}\lambda_m^{\alpha+\beta}(\lambda_m^{1-\alpha-\beta}|a_{nm}|^2+\lambda_n^{1-\alpha-\beta}|a_{mn}|^2)
$;

(ii) $|\mathrm{Corr}_{\rho}^{\alpha,\beta}|(A,B)
=\frac{1}{2}\sum_{mn}(\lambda_m-\lambda_m^{1-\beta}\lambda_n^{\beta})\overline{a_{nm}}b_{nm}
+(\lambda_m^{\alpha+\beta}\lambda_n^{1-\alpha-\beta}-\lambda_m^{\alpha}\lambda_n^{1-\alpha})a_{mn}\overline{b_{mn}}
=\frac{1}{2}\sum_{mn}\lambda_m^{\alpha}(\lambda_m^{\beta}-\lambda_n^{\beta})[\lambda_m^{1-\alpha-\beta}\overline{a_{nm}}b_{nm}+
\lambda_n^{1-\alpha-\beta}a_{mn}\overline{b_{mn}}]$;

(iii) $|\mathrm{I}_{\rho}^{\alpha,\beta}|(A)
=\frac{1}{2}\sum_{mn}\lambda_m^{\alpha}(\lambda_m^{\beta}-\lambda_n^{\beta})[\lambda_m^{1-\alpha-\beta}|a_{nm}|^2+
\lambda_n^{1-\alpha-\beta}|a_{mn}|^2]
=\frac{1}{2}\sum_{mn}(\lambda_m^{\alpha}-\lambda_n^{\alpha})(\lambda_m^{\beta}-\lambda_n^{\beta})\lambda_n^{1-\alpha-\beta}|a_{mn}|^2
=\frac{1}{2}\sum_{m<n}(\lambda_m^{\alpha}-\lambda_n^{\alpha})(\lambda_m^{\beta}-\lambda_n^{\beta})[\lambda_m^{1-\alpha-\beta}|a_{nm}|^2+
\lambda_n^{1-\alpha-\beta}|a_{mn}|^2] $;

(iv) $|\mathrm{J}_{\rho}^{\alpha,\beta}|(A)
=\frac{1}{2}\sum_{mn}\lambda_m^{\alpha}(\lambda_m^{\beta}+\lambda_n^{\beta})[\lambda_m^{1-\alpha-\beta}|a_{nm}|^2+
\lambda_n^{1-\alpha-\beta}|a_{mn}|^2]
=\frac{1}{2}\sum_{mn}(\lambda_m^{\alpha}+\lambda_n^{\alpha})(\lambda_m^{\beta}+\lambda_n^{\beta})\lambda_n^{1-\alpha-\beta}|a_{mn}|^2
\geq
\frac{1}{2}\sum_{m<n}(\lambda_m^{\alpha}+\lambda_n^{\alpha})(\lambda_m^{\beta}+\lambda_n^{\beta})[\lambda_m^{1-\alpha-\beta}|a_{nm}|^2+
\lambda_n^{1-\alpha-\beta}|a_{mn}|^2] $.

{\bf Proof.} Direct calculation shows that
$$\mathrm{Tr}(\rho A_0^\dag B_0)=\sum_{mn}\lambda_m \overline{a_{nm}}b_{nm},$$
$$\mathrm{Tr}(\rho^{\alpha+\beta} A_0^\dag \rho^{1-\alpha-\beta}B_0)=\sum_{mn}\lambda_m^{\alpha+\beta}\lambda_n^{1-\alpha-\beta}a_{mn}\overline{b_{mn}},$$
$$\mathrm{Tr}(\rho^{1-\beta} A_0^\dag \rho^{\beta}B_0)=\sum_{mn}\lambda_m^{1-\beta}\lambda_n^{\beta}\overline{a_{nm}}b_{nm},$$
$$\mathrm{Tr}(\rho^{\alpha} A_0^\dag \rho^{1-\alpha}B_0)=\sum_{mn}\lambda_m^{\alpha}\lambda_n^{1-\alpha}a_{mn}\overline{b_{mn}},$$
and we can thus obtain (i) and (ii) immediately. Consequently for (iii), we have
$$|\mathrm{I}_{\rho}^{\alpha,\beta}|(A)=|\mathrm{Corr}_{\rho}^{\alpha,\beta}|(A,A)
=\frac{1}{2}\sum_{mn}\lambda_m^{\alpha}(\lambda_m^{\beta}-\lambda_n^{\beta})[\lambda_m^{1-\alpha-\beta}|a_{nm}|^2+
\lambda_n^{1-\alpha-\beta}|a_{mn}|^2].$$ Moreover, we can rewrite
$|\mathrm{I}_{\rho}^{\alpha,\beta}|(A)$ as
$$|\mathrm{I}_{\rho}^{\alpha,\beta}|(A)
=\frac{1}{2}\sum_{mn}(\lambda_m^{\alpha}-\lambda_n^{\alpha})(\lambda_m^{\beta}-\lambda_n^{\beta})\lambda_n^{1-\alpha-\beta}|a_{mn}|^2$$
or
$$|\mathrm{I}_{\rho}^{\alpha,\beta}|(A)
=\frac{1}{2}\sum_{m<n}(\lambda_m^{\alpha}-\lambda_n^{\alpha})(\lambda_m^{\beta}-\lambda_n^{\beta})[\lambda_m^{1-\alpha-\beta}|a_{nm}|^2+      \lambda_n^{1-\alpha-\beta}|a_{mn}|^2].
$$
(iv) can be proved in a similar way. This completes the proof.
$\Box$

Now, we define the {\it modified weighted generalized Wigner-Yanase-Dyson} (MWGWYD) skew information  for an operator $A\in L^2(H)$ (not necessarily Hermitian) and $\rho\in D(H)$ as follows:
\begin{equation}\label{eq10}
|\mathrm{K}_{\rho}^{\alpha,\beta}|(A)=-\frac{1}{2}\mathrm{Tr}\left(\left[\frac{\rho^\alpha+\rho^\beta}{2},
A_0^\dag\right]\left[\frac{\rho^\alpha+\rho^\beta}{2},
A_0\right]\rho^{1-\alpha-\beta}\right),~~\alpha,\beta\geq
0,~\alpha+\beta\leq 1.
\end{equation}
A related quantity $|L_{\rho}^{\alpha,\beta}|(A)$ is defined as
\begin{equation}\label{eq11}
|\mathrm{L}_{\rho}^{\alpha,\beta}|(A)=\frac{1}{2}\mathrm{Tr}\left(\left\{\frac{\rho^\alpha+\rho^\beta}{2},
A_0^\dag\right\}\left\{\frac{\rho^\alpha+\rho^\beta}{2},
A_0\right\}\rho^{1-\alpha-\beta}\right),~~\alpha,\beta\geq
0,~\alpha+\beta\leq 1.
\end{equation}

The properties of the above two quantities are summarized in the following two propositions.

{\bf Proposition 4} Let $\alpha,\beta\geq 0$, $\alpha+\beta\leq 1$, $A\in L^2(H)$ and $\rho\in D(H)$. The following statements hold:

(i)
$|\mathrm{K}_{\rho}^{\alpha,\beta}|(A)=|\mathrm{K}_{\rho}^{\beta,\alpha}|(A)$;

(ii)
$|\mathrm{K}_{\rho}^{\alpha,\beta}|(A^\dag)=|\mathrm{K}_{\rho}^{\alpha,\beta}|(A)$;

(iii) $|\mathrm{K}_{\rho}^{\alpha,\beta}|(A)\geq
|\mathrm{I}_{\rho}^{\alpha,\beta}|(A)$.

{\bf Proof.} (i) and (ii) can be easily verified from the definition. We now prove (iii). By Proposition 1 (i), we have
\begin{eqnarray*}
|\mathrm{K}_{\rho}^{\alpha,\beta}|(A)
&=&-\frac{1}{2}\mathrm{Tr}([\frac{\rho^\alpha+\rho^\beta}{2},
A_0^\dag][\frac{\rho^\alpha+\rho^\beta}{2},
A_0]\rho^{1-\alpha-\beta})
\\&=&-\frac{1}{8}\mathrm{Tr}([\rho^\alpha, A_0^\dag][\rho^\alpha, A_0]\rho^{1-\alpha-\beta}+[\rho^\beta, A_0^\dag][\rho^\beta, A_0]\rho^{1-\alpha-\beta})+\frac{1}{4}(|\mathrm{I}_{\rho}^{\alpha,\beta}|(A)+|\mathrm{I}_{\rho}^{\beta,\alpha}|(A))
\\&=&-\frac{1}{8}\mathrm{Tr}([\rho^\alpha, A_0^\dag][\rho^\alpha, A_0]\rho^{1-\alpha-\beta}+[\rho^\beta, A_0^\dag][\rho^\beta, A_0]\rho^{1-\alpha-\beta})+\frac{1}{2}|\mathrm{I}_{\rho}^{\alpha,\beta}|(A).
\end{eqnarray*}
Suppose that the spectral decomposition of $\rho$ is $\rho=\sum_{m}\lambda_m |\psi_m\rangle\langle \psi_m|$ and denote $a_{mn}=\langle\psi_m|A_0|\psi_n \rangle$. Then we obtain
\begin{eqnarray*}
\mathrm{Tr}([\rho^\alpha, A_0^\dag][\rho^\alpha, A_0]\rho^{1-\alpha-\beta})
&=& \mathrm{Tr}(\rho^\alpha A_0^\dag-A_0^\dag\rho^\alpha)(\rho^\alpha A_0-A_0\rho^\alpha)\rho^{1-\alpha-\beta}
\\&=& \mathrm{Tr}(2\rho^\alpha A_0^\dag \rho^\alpha A_0-\rho^{2\alpha}A_0^\dag A_0-\rho^{2\alpha}A_0 A_0^\dag)\rho^{1-\alpha-\beta}
\\&=& \mathrm{Tr}(2\rho^{1-\beta} A_0^\dag \rho^\alpha A_0-\rho^{1+\alpha-\beta}A_0^\dag A_0-\rho^{1+\alpha-\beta}A_0 A_0^\dag)
\\&=& \sum_{mn}(2\lambda_n^{1-\beta}\lambda_m^{\alpha}-\lambda_n^{1+\alpha-\beta}-\lambda_m^{1+\alpha-\beta})|a_{mn}|^2,
\end{eqnarray*}
and
\begin{eqnarray*}
\mathrm{Tr}([\rho^\alpha, A_0^\dag][\rho^\alpha, A_0]\rho^{1-\alpha-\beta})
=\sum_{mn}(2\lambda_n^{1-\alpha}\lambda_m^{\beta}-\lambda_n^{1-\alpha+\beta}-\lambda_m^{1-\alpha+\beta})|a_{mn}|^2.
\end{eqnarray*}
By Proposition 3 (iii), we get
\begin{eqnarray*}
|\mathrm{K}_{\rho}^{\alpha,\beta}|(A)
&=&-\frac{1}{8}(\lambda_m^{1+\alpha-\beta}+\lambda_n^{1+\alpha-\beta}-2\lambda_n^{1-\beta}\lambda_m^{\alpha}
+\lambda_m^{1-\alpha+\beta}+\lambda_n^{1-\alpha+\beta}-2\lambda_n^{1-\alpha}\lambda_m^{\beta})
\\& &+\frac{1}{4}\sum_{mn}(\lambda_m^{\alpha}-\lambda_n^{\alpha})(\lambda_m^{\beta}-\lambda_n^{\beta})\lambda_n^{1-\alpha-\beta}|a_{mn}|^2.
\end{eqnarray*}
Since $\alpha,\beta\geq 0$, $\alpha+\beta\leq 1$ and $0\leq \lambda_m,\lambda_n \leq 1$, we have $\lambda_m^{1+\alpha-\beta}\geq \lambda_m^{2\alpha}$, $\lambda_m^{1-\alpha+\beta}\geq \lambda_m^{2\beta}$ and $\lambda_n^{1-\alpha-\beta}\leq 1$, and thus
$$\lambda_m^{1+\alpha-\beta}+\lambda_m^{1+\alpha-\beta}\geq (\lambda_m^{2\alpha}+\lambda_m^{2\beta})\lambda_n^{1-\alpha-\beta},$$
which implies that
\begin{eqnarray*}
&&\lambda_m^{1+\alpha-\beta}+\lambda_n^{1+\alpha-\beta}-2\lambda_n^{1-\beta}\lambda_m^{\alpha}
+\lambda_m^{1-\alpha+\beta}+\lambda_n^{1-\alpha+\beta}-2\lambda_n^{1-\alpha}\lambda_m^{\beta}
\\&&\geq  \lambda_m^{2\alpha}\lambda_n^{1-\alpha-\beta}+\lambda_n^{1+\alpha-\beta}-2\lambda_n^{1-\beta}\lambda_m^{\alpha}
+\lambda_m^{2\beta}\lambda_n^{1-\alpha-\beta}+\lambda_n^{1-\alpha+\beta}-2\lambda_n^{1-\alpha}\lambda_m^{\beta}
\\&&=[(\lambda_m^{\alpha}-\lambda_n^{\alpha})^2+(\lambda_m^{\beta}-\lambda_n^{\beta})^2]\lambda_n^{1-\alpha-\beta}
\\&&\geq 2(\lambda_m^{\alpha}-\lambda_n^{\alpha})(\lambda_m^{\beta}-\lambda_n^{\beta})\lambda_n^{1-\alpha-\beta},
\end{eqnarray*}
Again, by Proposition 3 (iii), we conclude that
$|\mathrm{K}_{\rho}^{\alpha,\beta}|(A)\geq
|\mathrm{I}_{\rho}^{\alpha,\beta}|(A)$. $\Box$

In a similar way, we can prove the following proposition.

{\bf Proposition 5} Let $\alpha,\beta\geq 0$, $\alpha+\beta\leq 1$, $A\in L^2(H)$ and $\rho\in D(H)$. The following statements hold:

(i)
$|\mathrm{L}_{\rho}^{\alpha,\beta}|(A)=|\mathrm{L}_{\rho}^{\beta,\alpha}|(A)$;

(ii)
$|\mathrm{L}_{\rho}^{\alpha,\beta}|(A^\dag)=|\mathrm{L}_{\rho}^{\alpha,\beta}|(A)$;

(iii) $|\mathrm{L}_{\rho}^{\alpha,\beta}|(A)\geq
|\mathrm{J}_{\rho}^{\alpha,\beta}|(A)$.

\vskip0.1in

\noindent {\bf 3. Uncertainty relations based on MGWYD and MWGWYD skew information}

In this section, we present some new uncertainty relations based on
MGWYD skew information and MWGWYD skew information and related
quantities defined in the previous section. First of all, imitating
the proof of Lemma 1 and Lemma 2 in \cite{FYJ}, we can prove the
following lemma.

{\bf Lemma 1} For any $x,y\geq 0$, $0\leq \beta\leq
\min\{\alpha,1-\alpha\}$, we have
\begin{equation}\label{eq12}
(x^\alpha+y^\alpha)|x^\beta-y^\beta|\leq |x-y|,
\end{equation}
\begin{equation}\label{eq13}
4\alpha\beta(x-y)^2\leq (x^{2\alpha}-y^{2\alpha})(x^{2\beta}-y^{2\beta}).
\end{equation}

Utilizing Lemma 1, the first main result of this paper can be stated
as follows.

{\bf Theorem 1} Let $0\leq \beta\leq \min\{\alpha,1-\alpha\}$,
$A,B\in L^2(H)$ and $\rho\in D(H)$. We have
\begin{equation}\label{eq14}
|\mathrm{U}_{\rho}^{\alpha,\beta}|(A)\cdot
|\mathrm{U}_{\rho}^{\alpha,\beta}|(B)\geq
4\alpha\beta||\mathrm{Corr}_{\rho}^{\alpha,\beta}|(A,B)|^2.
\end{equation}

{\bf Proof.} It follows from Proposition 3 (ii) that
\begin{eqnarray*}
|\mathrm{Corr}_{\rho}^{\alpha,\beta}|(A,B)
&=&\frac{1}{2}\sum_{m<n}\lambda_m^{\alpha}(\lambda_m^{\beta}-\lambda_n^{\beta})[\lambda_m^{1-\alpha-\beta}\overline{a_{nm}}b_{nm}+
\lambda_n^{1-\alpha-\beta}a_{mn}\overline{b_{mn}}]\notag
\\&&+\frac{1}{2}\sum_{m<n}\lambda_n^{\alpha}(\lambda_n^{\beta}-\lambda_m^{\beta})[\lambda_n^{1-\alpha-\beta}\overline{a_{mn}}b_{mn}+      \lambda_n^{1-\alpha-\beta}a_{nm}\overline{b_{nm}}],
\end{eqnarray*}
which implies that
\begin{eqnarray*}
||\mathrm{Corr}_{\rho}^{\alpha,\beta}|(A,B)| &\leq&
\frac{1}{2}\sum_{m<n}\lambda_m^{\alpha}|\lambda_m^{\beta}-\lambda_n^{\beta}|\cdot|\lambda_m^{1-\alpha-\beta}\overline{a_{nm}}b_{nm}+
\lambda_n^{1-\alpha-\beta}a_{mn}\overline{b_{mn}}|\notag
\\&&+\frac{1}{2}\sum_{m<n}\lambda_n^{\alpha}|\lambda_n^{\beta}-\lambda_m^{\beta}|\cdot|\lambda_n^{1-\alpha-\beta}\overline{a_{mn}}b_{mn}+      \lambda_n^{1-\alpha-\beta}a_{nm}\overline{b_{nm}}|
\\&=&\frac{1}{2}\sum_{m<n}(\lambda_m^{\alpha}+\lambda_n^{\alpha})|\lambda_m^{\beta}-\lambda_n^{\beta}|     \cdot|\lambda_m^{1-\alpha-\beta}\overline{a_{nm}}b_{nm}+\lambda_n^{1-\alpha-\beta}a_{mn}\overline{b_{mn}}|.
\end{eqnarray*}
Utilizing (\ref{eq12}), we obtain
\begin{equation}\label{eq15}
||\mathrm{Corr}_{\rho}^{\alpha,\beta}|(A,B)|\leq
\frac{1}{2}\sum_{m<n}|\lambda_m-\lambda_n|
\cdot|\lambda_m^{1-\alpha-\beta}\overline{a_{nm}}b_{nm}+\lambda_n^{1-\alpha-\beta}a_{mn}\overline{b_{mn}}|.
\end{equation}
Thus, combining Cauchy-Schwarz inequality, Lemma 1 (\ref{eq13}), Proposition (iii) and (iv), from Eq. (\ref{eq15}) we obtain
\begin{eqnarray*}
&&4\alpha\beta||\mathrm{Corr}_{\rho}^{\alpha,\beta}|(A,B)|^2
\\&\leq& \alpha\beta[\sum_{m<n}|\lambda_m-\lambda_n|\cdot
|\lambda_m^{1-\alpha-\beta}\overline{a_{nm}}b_{nm}+\lambda_n^{1-\alpha-\beta}a_{mn}\overline{b_{mn}}|]^2
\\&=&\frac{1}{4}[\sum_{m<n}2\sqrt{\alpha\beta}|\lambda_m-\lambda_n|\cdot
|\lambda_m^{1-\alpha-\beta}\overline{a_{nm}}b_{nm}+\lambda_n^{1-\alpha-\beta}a_{mn}\overline{b_{mn}}|]^2
\\&\leq&\frac{1}{4}[\sum_{m<n}[(\lambda_m^{\alpha}-\lambda_n^{\alpha})(\lambda_m^{\beta}-\lambda_n^{\beta})
(\lambda_m^{\alpha}+\lambda_n^{\alpha})(\lambda_m^{\beta}+\lambda_n^{\beta})]^{\frac{1}{2}}\cdot
|\lambda_m^{1-\alpha-\beta}\overline{a_{nm}}b_{nm}+\lambda_n^{1-\alpha-\beta}a_{mn}\overline{b_{mn}}|]^2
\\&\leq&\frac{1}{4}[\sum_{m<n}[(\lambda_m^{\alpha}-\lambda_n^{\alpha})(\lambda_m^{\beta}-\lambda_n^{\beta})
(\lambda_m^{1-\alpha-\beta}|a_{nm}|^2+\lambda_n^{1-\alpha-\beta}|a_{mn}|^2)]^{\frac{1}{2}}\cdot
\notag
\\&& [(\lambda_m^{\alpha}+\lambda_n^{\alpha})(\lambda_m^{\beta}+\lambda_n^{\beta})
(\lambda_m^{1-\alpha-\beta}|b_{nm}|^2+\lambda_n^{1-\alpha-\beta}|b_{mn}|^2)]^{\frac{1}{2}}]^2
\\&\leq& \frac{1}{4}\sum_{m<n}(\lambda_m^{\alpha}-\lambda_n^{\alpha})(\lambda_m^{\beta}-\lambda_n^{\beta})
(\lambda_m^{1-\alpha-\beta}|a_{nm}|^2+\lambda_n^{1-\alpha-\beta}|a_{mn}|^2)\cdot
\notag
\\&& \sum_{m<n}(\lambda_m^{\alpha}+\lambda_n^{\alpha})(\lambda_m^{\beta}+\lambda_n^{\beta})
(\lambda_m^{1-\alpha-\beta}|b_{nm}|^2+\lambda_n^{1-\alpha-\beta}|b_{mn}|^2)
\\&\leq& |\mathrm{I}_{\rho}^{\alpha,\beta}|(A)\cdot |\mathrm{J}_{\rho}^{\alpha,\beta}|(B).
\end{eqnarray*}
From the above deductions, we can also obtain
$$4\alpha\beta||\mathrm{Corr}_{\rho}^{\alpha,\beta}|(A,B)|^2
\leq |\mathrm{I}_{\rho}^{\alpha,\beta}|(B)\cdot
|\mathrm{J}_{\rho}^{\alpha,\beta}|(A).
$$
Therefore, by Proposition 2 (iv), we get
$$
|\mathrm{U}_{\rho}^{\alpha,\beta}|(A)\cdot
|\mathrm{U}_{\rho}^{\alpha,\beta}|(B)\geq
4\alpha\beta||\mathrm{Corr}_{\rho}^{\alpha,\beta}|(A,B)|^2.
$$
This completes the proof.
$\Box$

Imitating the proof of Lemma 3 in  \cite{FYJ}, we can prove the following lemma.

{\bf Lemma 2}  For any $x,y\geq 0$, $0\leq \beta\leq
\min\{4\alpha,1-\alpha\}$, we have
\begin{equation}\label{eq16}
(x^{\alpha+\beta}-x^{\alpha}y^{\beta})^2\leq (x^{2\alpha}-y^{2\alpha})(x^{2\beta}-y^{2\beta}).
\end{equation}

Based on this lemma, we now give the second main result of this paper.

{\bf Theorem 2} Let $0\leq \beta\leq \min\{4\alpha,1-\alpha\}$,
$A,B\in L^2(H)$ and $\rho\in D(H)$. Then we have
\begin{equation}\label{eq17}
|\mathrm{U}_{\rho}^{\alpha,\beta}|(A)\cdot
|\mathrm{U}_{\rho}^{\alpha,\beta}|(B)\geq
\frac{1}{4}||\mathrm{Corr}_{\rho}^{\alpha,\beta}|(A,B)|^2.
\end{equation}

{\bf Proof.} It follows from Proposition 3 (ii), (iii), (iv), Lemma 2 and Cauchy-Schwarz inequality that
\begin{eqnarray*}
&&||\mathrm{Corr}_{\rho}^{\alpha,\beta}|(A,B)|^2
\\&=& \frac{1}{4}|\sum_{mn}\lambda_m^{\alpha}(\lambda_m^{\beta}-\lambda_n^{\beta})[\lambda_m^{1-\alpha-\beta}\overline{a_{nm}}b_{nm}+      \lambda_n^{1-\alpha-\beta}a_{mn}\overline{b_{mn}}]|^2
\\&\leq& \frac{1}{4}[\sum_{mn}\lambda_m^{\alpha}|\lambda_m^{\beta}-\lambda_n^{\beta}|\cdot|\lambda_m^{1-\alpha-\beta}\overline{a_{nm}}b_{nm}+      \lambda_n^{1-\alpha-\beta}a_{mn}\overline{b_{mn}}]^2
\\&\leq& \frac{1}{4}[\sum_{mn}[(\lambda_m^{\alpha}-\lambda_n^{\alpha})(\lambda_m^{\beta}-\lambda_n^{\beta})
(\lambda_m^{\alpha}+\lambda_n^{\alpha})(\lambda_m^{\beta}+\lambda_n^{\beta})]^{\frac{1}{2}}\cdot\notag
\\& &(\lambda_m^{1-\alpha-\beta}|a_{nm}|^2+\lambda_n^{1-\alpha-\beta}|a_{mn}|^2)^{\frac{1}{2}}
(\lambda_m^{1-\alpha-\beta}|b_{nm}|^2+\lambda_n^{1-\alpha-\beta}|b_{mn}|^2)^{\frac{1}{2}}]^2
\\&\leq& \frac{1}{4}[\sum_{mn}[(\lambda_m^{\alpha}-\lambda_n^{\alpha})(\lambda_m^{\beta}-\lambda_n^{\beta})
(\lambda_m^{\alpha}+\lambda_n^{\alpha})(\lambda_m^{\beta}+\lambda_n^{\beta})]^{\frac{1}{2}}\cdot\notag
\\& &(\lambda_m^{1-\alpha-\beta}|a_{nm}|^2+\lambda_n^{1-\alpha-\beta}|a_{mn}|^2)^{\frac{1}{2}}
(\lambda_m^{1-\alpha-\beta}|b_{nm}|^2+\lambda_n^{1-\alpha-\beta}|b_{mn}|^2)^{\frac{1}{2}}]^2
\\&\leq& \frac{1}{4}\sum_{mn}(\lambda_m^{\alpha}-\lambda_n^{\alpha})(\lambda_m^{\beta}-\lambda_n^{\beta})
(\lambda_m^{1-\alpha-\beta}|a_{nm}|^2+\lambda_n^{1-\alpha-\beta}|a_{mn}|^2)\cdot
\notag
\\&& \sum_{mn}(\lambda_m^{\alpha}+\lambda_n^{\alpha})(\lambda_m^{\beta}+\lambda_n^{\beta})
(\lambda_m^{1-\alpha-\beta}|b_{nm}|^2+\lambda_n^{1-\alpha-\beta}|b_{mn}|^2)
\\&=& \frac{1}{4}\cdot 4|\mathrm{I}_{\rho}^{\alpha,\beta}|(A)\cdot 4|\mathrm{J}_{\rho}^{\alpha,\beta}|(B)
\\&=& 4|\mathrm{I}_{\rho}^{\alpha,\beta}|(A)\cdot |\mathrm{J}_{\rho}^{\alpha,\beta}|(B).
\end{eqnarray*}
Similarly, we have
$$
||\mathrm{Corr}_{\rho}^{\alpha,\beta}|(A,B)|^2\leq
4|\mathrm{I}_{\rho}^{\alpha,\beta}|(B)\cdot
|\mathrm{J}_{\rho}^{\alpha,\beta}|(A).
$$
Hence, by Proposition 2 (iv), we conclude that
$$
||\mathrm{Corr}_{\rho}^{\alpha,\beta}|(A,B)|^2\leq
4|\mathrm{U}_{\rho}^{\alpha,\beta}|(A)\cdot
|\mathrm{U}_{\rho}^{\alpha,\beta}|(B).
$$
This completes the proof.
$\Box$

{\bf Remark} Let us compare the above two theorems. (\ref{eq17})
holds when $\alpha,\beta\geq 0$, $\alpha+\beta\leq 1$ and
$\alpha\leq \beta\leq 4\alpha$. When $\alpha,\beta\geq 0$,
$\alpha+\beta\leq 1$ and $\frac{1}{4}\leq \beta\leq \alpha$, we have
$4\alpha\beta\geq \frac{1}{4}$, and thus (\ref{eq17}) is better than
(\ref{eq14}). When $\alpha,\beta\geq 0$, $\alpha+\beta\leq 1$ and
$\beta\leq \alpha\leq \frac{1}{4}$, we have $4\alpha\beta\leq
\frac{1}{4}$ and thus (\ref{eq14}) is better than (\ref{eq17}).

In the previous section, we have defined the quantities
$|\mathrm{K}_{\rho}^{\alpha,\beta}|(A)$ and
$|\mathrm{L}_{\rho}^{\alpha,\beta}|(A)$. Now we define the quantity
$|\mathrm{W}_{\rho}^{\alpha,\beta}|(A)=\sqrt{|\mathrm{K}_{\rho}^{\alpha,\beta}|(A)|\mathrm{L}_{\rho}^{\alpha,\beta}|(A)}$
for any $A\in L^2(H)$. Then it follows from Proposition 2 (iv),
Proposition 4 (iii) and Proposition 5 (iii) that
$$|\mathrm{W}_{\rho}^{\alpha,\beta}|(A)\geq |\mathrm{U}_{\rho}^{\alpha,\beta}|(A), |\mathrm{W}_{\rho}^{\alpha,\beta}|(B)\geq |\mathrm{U}_{\rho}^{\alpha,\beta}|(B),\,\,\makebox{for all}\,\, A,B\in L^2(H).$$
Therefore, we obtain the following two uncertainty relations as
consequences of Theorem 1 and Theorem 2.

{\bf Corollary 1} Let $0\leq \beta\leq \min\{\alpha,1-\alpha\}$,
$A,B\in L^2(H)$ and $\rho\in D(H)$. Then we have
\begin{equation}\label{eq18}
|\mathrm{W}_{\rho}^{\alpha,\beta}|(A)\cdot
|\mathrm{W}_{\rho}^{\alpha,\beta}|(B)\geq
4\alpha\beta||\mathrm{Corr}_{\rho}^{\alpha,\beta}|(A,B)|^2.
\end{equation}

{\bf Corollary 2} Let $0\leq \beta\leq \min\{4\alpha,1-\alpha\}$,
$A,B\in L^2(H)$ and $\rho\in D(H)$. Then we have
\begin{equation}\label{eq19}
|\mathrm{W}_{\rho}^{\alpha,\beta}|(A)\cdot
|\mathrm{W}_{\rho}^{\alpha,\beta}|(B)\geq
\frac{1}{4}||\mathrm{Corr}_{\rho}^{\alpha,\beta}|(A,B)|^2.
\end{equation}

{\bf Example 1} Consider the Werner state
$$\rho_w^{ab}=\left(\begin{array}{cccc}
         \frac{1}{3}p&0&0&0\\
         0&\frac{1}{6}(3-2p)&\frac{1}{6}(4p-3)&0\\
         0&\frac{1}{6}(4p-3)&\frac{1}{6}(3-2p)&0\\
         0&0&0&\frac{1}{3}p\\
         \end{array}
         \right),
$$
where $p\in [0,1]$. Note that $\rho_w^{ab}$ is separable when $p\in
[0,\frac{1}{3}]$. Let $A$ and $B$ be the following non-Hermitian matrices
\begin{equation}\label{e1}
A=\left(
\begin{array}{cccc}
 0 & 1 & 0 & -i \\
 1 & 0 & i & 0 \\
 1 & 0 & 1 & 0 \\
 0 & -1 & 0 & 1 \\
\end{array}
\right),~~~
B=\left(
\begin{array}{cccc}
 1 & 0 & 1 & 0 \\
 0 & 1 & 0 & -1 \\
 0 & 1 & 0 & -i \\
 1 & 0 & i & 0 \\
\end{array}
\right).
\end{equation}
Figure 1 illustrates the uncertainty relations of
Eq. (\ref{eq14}) with different values of $\alpha$ and $\beta$.

\begin{figure}[ht]\centering
{\begin{minipage}[b]{0.6\linewidth}
\includegraphics[width=0.8\textwidth]{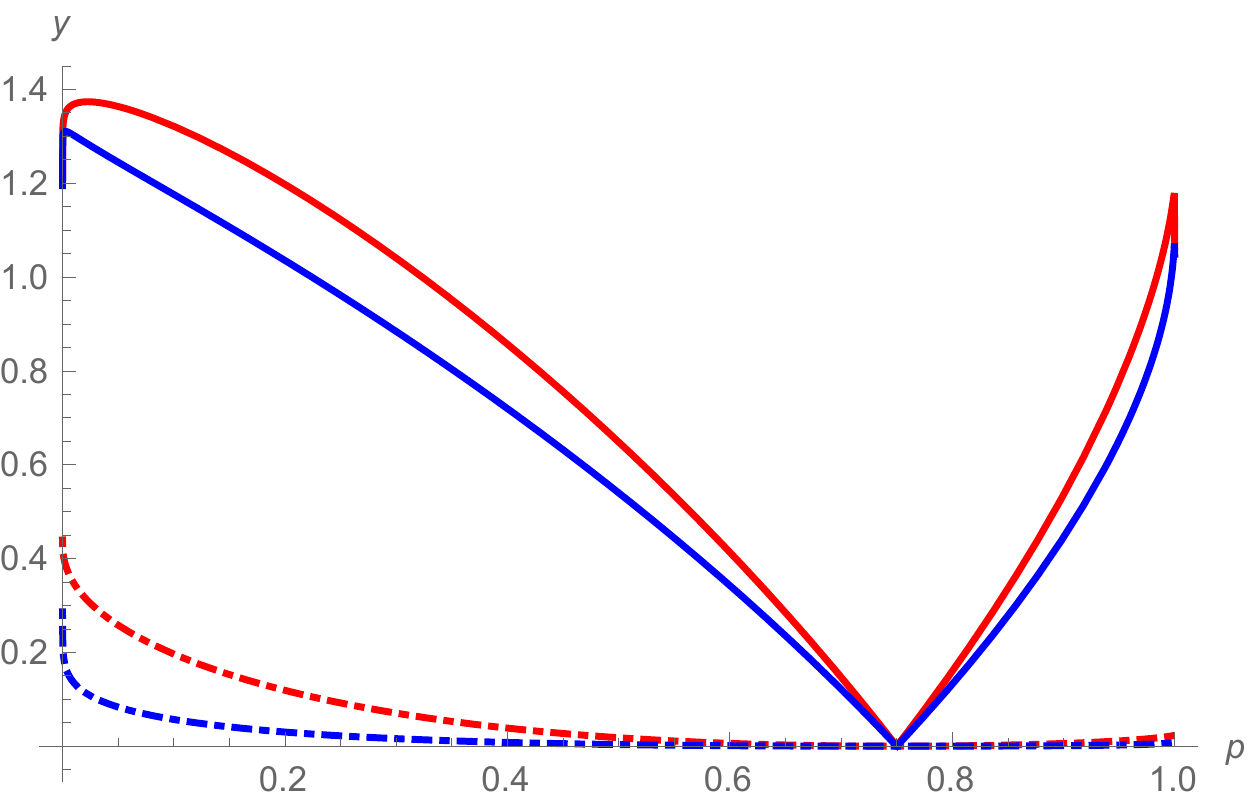}
\end{minipage}}
\caption{The $y$-axis shows the uncertainty and its lower bounds.
Red solid (dotdashed) line represents the value of the left
(right)-hand side of Eq. (\ref{eq14}) with $\alpha=\frac{11}{20}$
and $\beta=\frac{2}{5}$ for $\rho_w^{ab}$; blue solid (dotdashed)
line represents the value of the left (right)-hand side of Eq.
(\ref{eq14}) with $\alpha=\frac{15}{20}$ and $\beta=\frac{1}{5}$ for
$\rho_w^{ab}$.} \label{fig:u1}
\end{figure}

Moreover, when we fix the value of $p$, the gap between the left and
right hand sides of Eq. (\ref{eq14}) for separable states are
greater than those for the entangled states. See Figure 2 for an
illustration of this fact for $p=0.3$ and $p=0.9$.
\begin{figure}[ht]\centering
\subfigure[] {\begin{minipage}[b]{0.47\linewidth}
\includegraphics[width=1\textwidth]{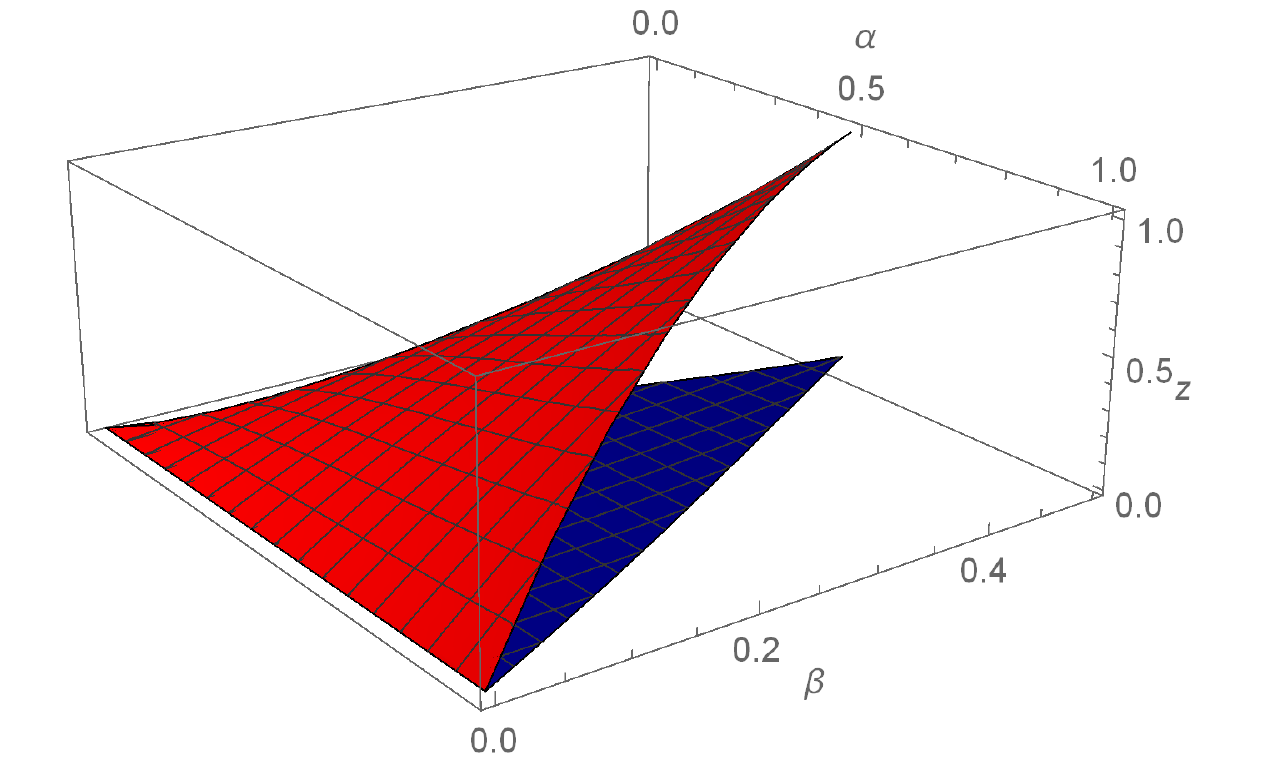}
\end{minipage}}
\subfigure[] {\begin{minipage}[b]{0.47\linewidth}
\includegraphics[width=1\textwidth]{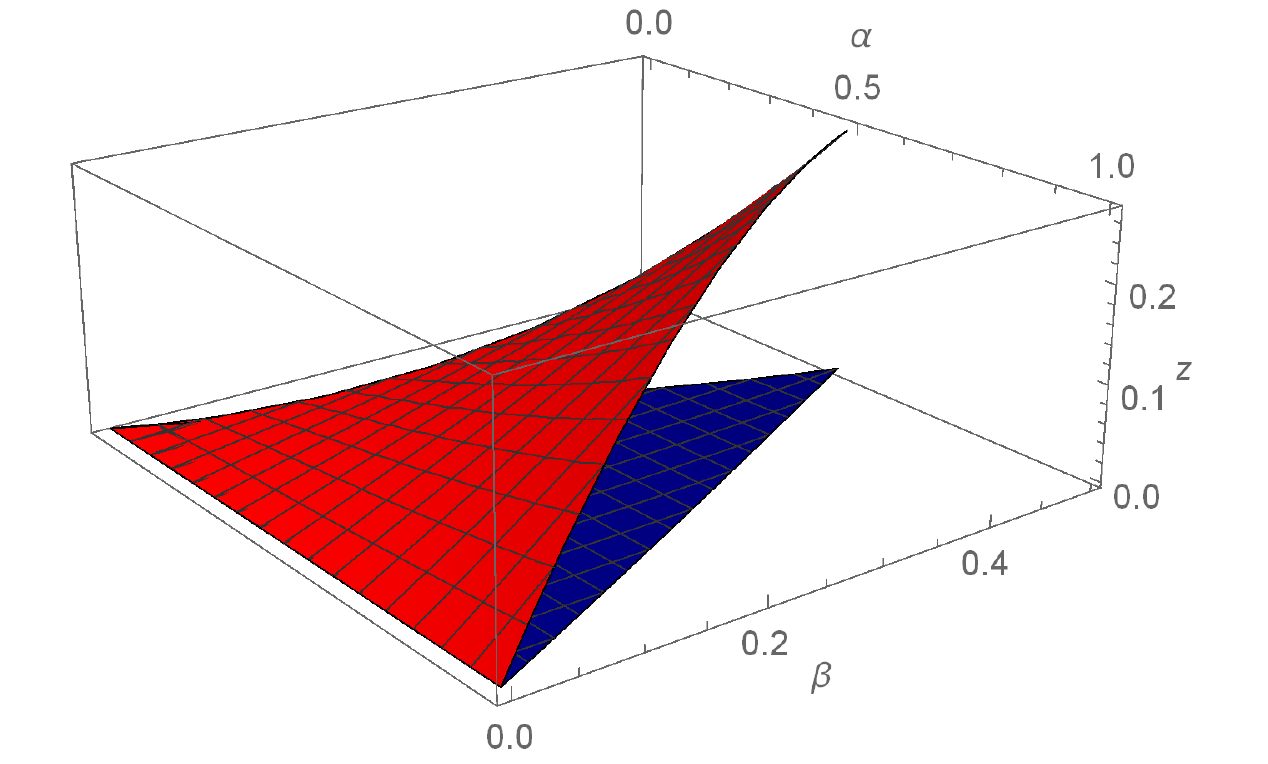}
\end{minipage}}
\caption{The $z$-axis shows the uncertainty and its lower bounds.
(a) $p=0.3$ (in this case $\rho_{w}^{ab}$ is a separable state): Red
(blue) surface represents the value of the left (right)-hand side of
Eq. (\ref{eq14}) for $\rho_w^{ab}$; (b) $p=0.9$ (in this case
$\rho_{w}^{ab}$ is an entangled state): Red (blue) surface
represents the value of the left (right)-hand side of Eq.
(\ref{eq14}) for $\rho_w^{ab}$.} \label{fig:u2u3}
\end{figure}

{\bf Example 2} Consider the isotropic state
$$\rho_{iso}^{ab}=\left(
\begin{array}{cccc}
 \frac{1}{6} (2 F+1) & 0 & 0 & \frac{1}{6} (4 F-1) \\
 0 & \frac{1}{3}(1-F) & 0 & 0 \\
 0 & 0 & \frac{1}{3}(1-F) & 0 \\
 \frac{1}{6} (4 F-1) & 0 & 0 & \frac{1}{6} (2 F+1) \\
\end{array}
\right),
$$
where $F\in [0,1]$. Note that $\rho_{iso}^{ab}$ is separable when
$F\in [0,\frac{1}{2}]$. With $A$ and $B$ being the non-Hermitian
matrices given in (\ref{e1}), Figure 3 illustrates the uncertainty
relations of Eq. (\ref{eq14}) with different values of $\alpha$ and
$\beta$.

\begin{figure}[ht]\centering
{\begin{minipage}[b]{0.6\linewidth}
\includegraphics[width=0.8\textwidth]{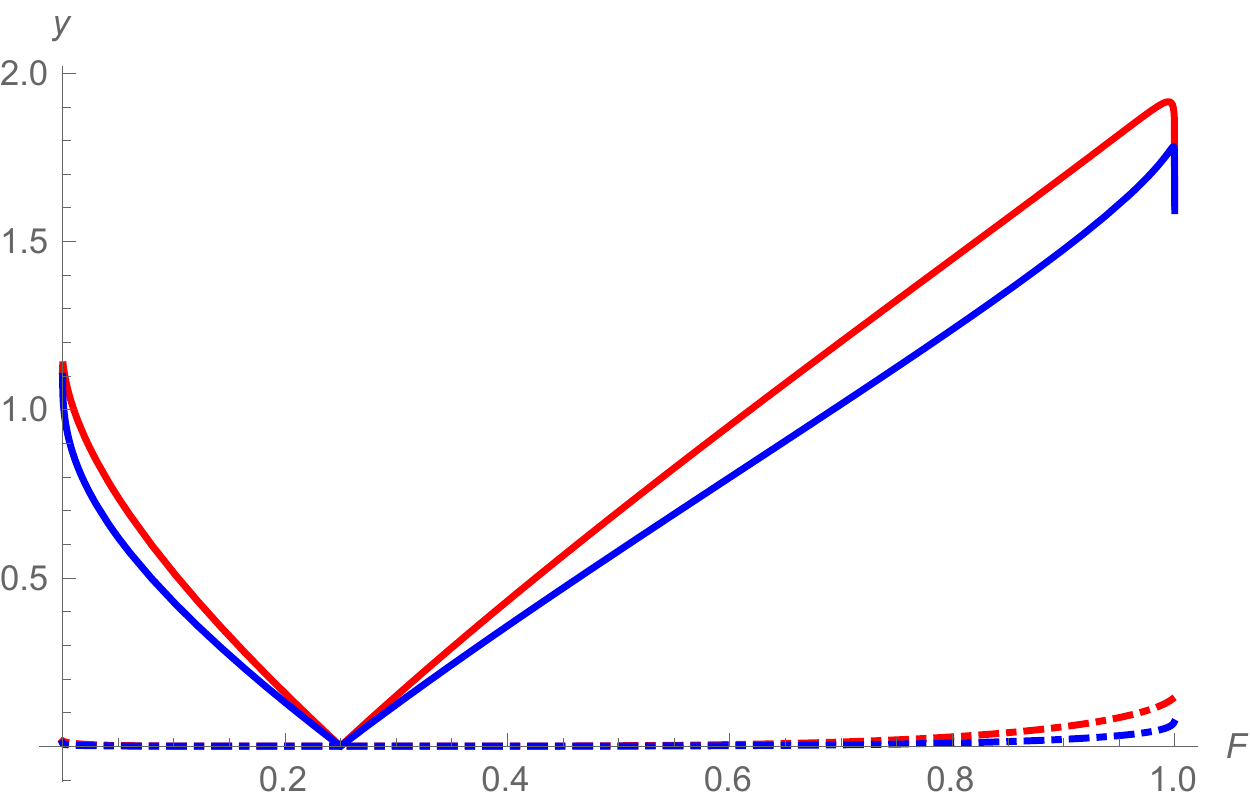}
\end{minipage}}
\caption{The $y$-axis shows the uncertainty and its lower bounds.
Red solid (dotdashed) line represents the value of the left
(right)-hand side of Eq. (\ref{eq14}) with $\alpha=\frac{11}{20}$
and $\beta=\frac{2}{5}$ for $\rho_{iso}^{ab}$; blue solid
(dotdashed) line represents the value of the left (right)-hand side
of Eq. (\ref{eq14}) with $\alpha=\frac{15}{20}$ and
$\beta=\frac{1}{5}$ for $\rho_{iso}^{ab}$.} \label{fig:u4}
\end{figure}

Moreover, when we fix the value of $F$, the gap between the left and
right hand sides of Eq. (\ref{eq14}) for separable states are less
than those for the entangled states. See Figure 4 for an
illustration of this fact for $F=0.4$ and $F=0.7$.
\begin{figure}[ht]\centering
\subfigure[] {\begin{minipage}[b]{0.47\linewidth}
\includegraphics[width=1\textwidth]{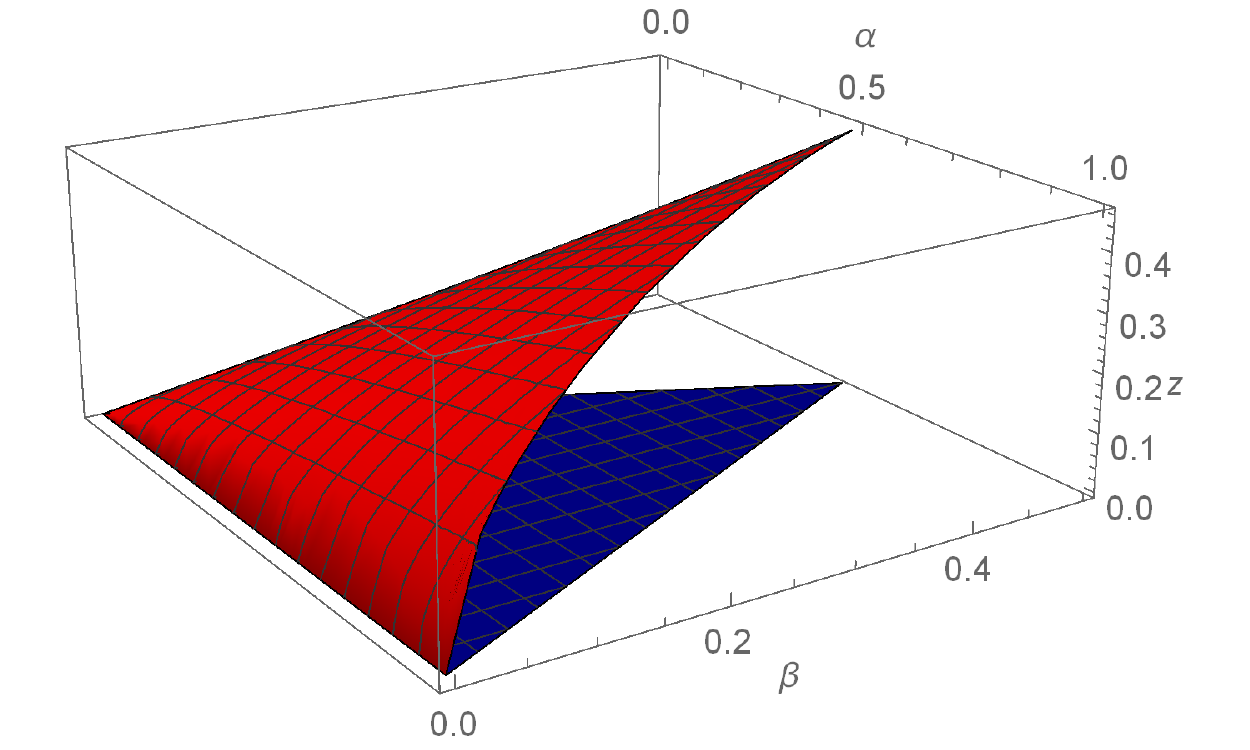}
\end{minipage}}
\subfigure[] {\begin{minipage}[b]{0.47\linewidth}
\includegraphics[width=1\textwidth]{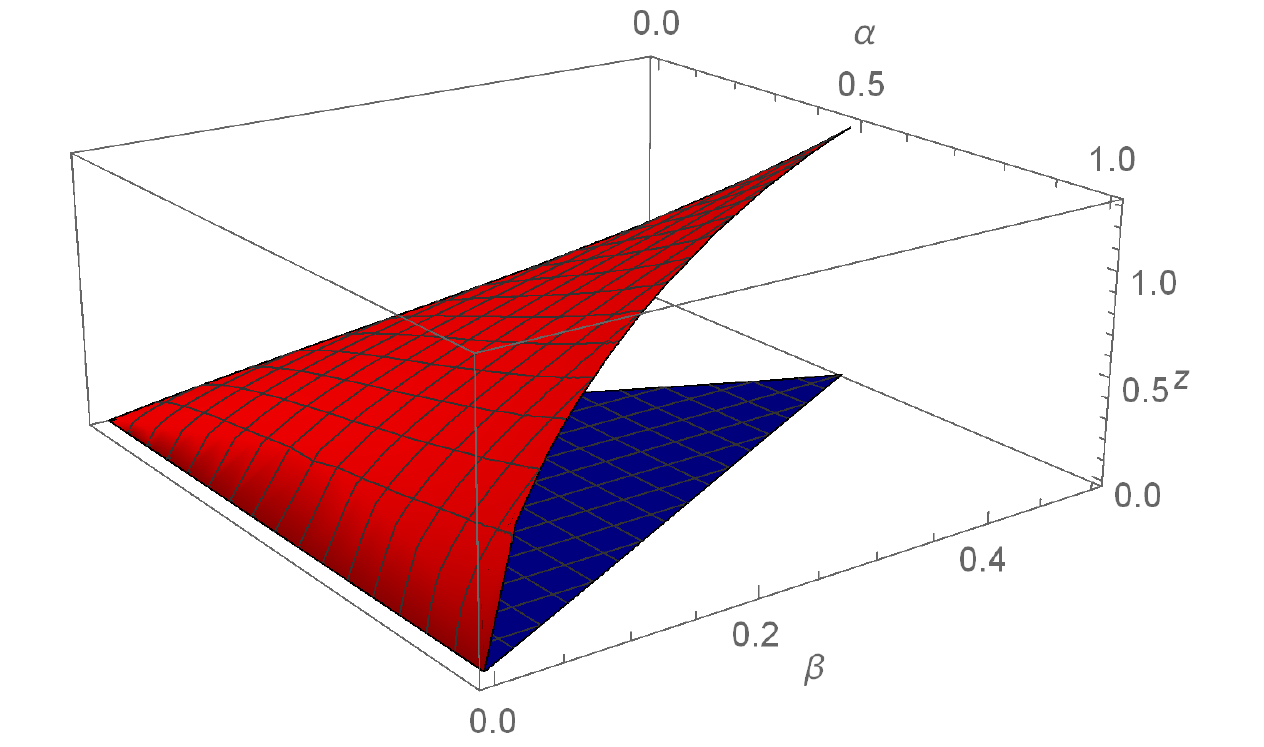}
\end{minipage}}
\caption{The $z$-axis shows the uncertainty and its lower bounds.
(a) $F=0.4$ (in this case $\rho_{iso}^{ab}$ is a separable state):
Red (blue) surface represents the value of the left (right)-hand
side of Eq. (\ref{eq14}) for $\rho_{iso}^{ab}$; (b) $F=0.7$ (in this
case $\rho_{iso}^{ab}$ is an entangled state): Red (blue) surface
represents the value of the left (right)-hand side of Eq.
(\ref{eq14}) for $\rho_{iso}^{ab}$.} \label{fig:u5u6}
\end{figure}

\vskip0.1in

\noindent {\bf 4. Conclusions}

Based on the newly introduced quantities termed modified generalized Wigner-Yanase-Dyson skew information and modified weighted generalized Wigner-Yanase-Dyson skew information,
we have derived new uncertainty relations, which turned out to be the generalizations of the main results in \cite{FYJ}.
Information based quantum uncertainty relations are of significance for usual Hermitian quantum mechanical systems.
Our work shew new light on the study of uncertainty relations for non-Hermitian operators.

\vskip0.1in

\noindent

\subsubsection*{Acknowledgements}
We would like to thank Dr. Bing Yu for useful discussions. This work
was supported by the National Natural Science Foundation of China
(11701259, 11971140, 11461045, 11675113, 11875034, 11505091), China
Scholarship Council (201806825038), the Key Project of Beijing
Municipal Commission of Education (KZ201810028042), Beijing Natural
Science Foundation (Z190005), and the Major Program of Jiangxi
Provincial NSF (20161ACB21006). This work was completed while Zhaoqi
Wu was visiting the Max-Planck-Institute for Mathematics in the
Sciences in Germany.


\end{document}